\documentclass[twocolumn,showpacs,preprintnumbers,amsmath,amssymb,superscriptaddress]{revtex4}
\usepackage{graphicx}
\usepackage{dcolumn}
\usepackage{bm}
\usepackage{color}

\def\dca{$\approx$}
\def\ca{$\sim$}
\def\teod{TeO$_2$~}
\def\tect{$^{130}$Te }
\def\ciccio{5$\times$5$\times$5 cm$^3$ }
\def\magro{3$\times$3$\times$6 cm$^3$ }
\def\bb{$\beta\beta$ }
\def\per{$\times$}

\def\amnu{$\vert\langle m_{\nu} \rangle\vert$ }
\def\pom{$\pm$ }

\begin{document}


\title{A New Limit on the  neutrinoless \bb-decay of \tect}

\newcommand{\mib}{Dipartimento di Fisica dell'Universit\`{a} di 
Milano-Bicocca e Sezione di Milano dell'INFN, Milan I-20126,
Italy} 
\newcommand{\usc}{Department of Physics and Astronomy , University of South Carolina, Columbia , South
Carolina,29208,USA}
\newcommand{\lngs}{Laboratori Nazionali del Gran Sasso, I-67010,
Assergi (L'Aquila), Italy}
\newcommand{\ufi}{Dipartimento di Fisica dell'Universit\`{a} di 
Firenze  e Sezione di Firenze dell'INFN, Firenze  I-50125,
Italy} 
\newcommand{\lbnl}{Lawrence Berkeley National Laboratory , Berkeley, California 94720, USA} 
\newcommand{\ucb}{Dept. of Materials Science and Mineral
Engineering, University of California, Berkeley, California
94720, USA}
\newcommand{\co}{Dipartimento di Fisica e
Matematica dell'Universit\`{a} dell'Insubria e Sezione di Milano
 dell'INFN, Como I-22100, Italy }
\newcommand{\uge}{Dipartimento di Fisica dell'Universit\`{a} di
Genova e Sezione di Genova dell'INFN, Genova I-16146, Italy}
\newcommand{\uz}{Laboratorio de Fisica Nuclear y Alta Energias, Universitad de Zaragoza, 5001 Zaragoza , Spain} 
\newcommand{\ul}{Kamerling Onnes Laboratory, Leiden University,  2300
RAQ,Leiden, the Netherlands}
\newcommand{\lnl}{Laboratori Nazionali di Legnaro, Via Romea 4. I-35020 Legnaro,(padova),Italy}

\affiliation{\mib}
\affiliation{\usc}
\affiliation{\lngs}
\affiliation{\ufi}
\affiliation{\lbnl}
\affiliation{\uz}
\affiliation{\ul}
\affiliation{\uge}
\affiliation{\co}
\affiliation{\ucb}
\affiliation{\lnl}

\author{C.~Arnaboldi}\affiliation{\mib}
\author{D.R.~Artusa}\affiliation{\usc}
\author{F.T.~Avignone III}\affiliation{\usc}
\author{M.~Balata}\affiliation{\lngs}
\author{I.~Bandac}\affiliation{\usc}
\author{M.~Barucci}\affiliation{\ufi}
\author{J.W.~Beeman}\affiliation{\lbnl}
\author{C.~Brofferio}\affiliation{\mib}
\author{C.~Bucci}\affiliation{\lngs}
\author{S.~Capelli}\affiliation{\mib}
\author{L.~Carbone}\affiliation{\mib}
\author{S.~Cebrian}, \affiliation{\uz}
\author{O.~Cremonesi}\affiliation{\mib}
\author{R.J.Creswick}\affiliation{\usc}
\author{A.~de Waard}\affiliation{\ul}
\author{H.A.Farach}\affiliation{\usc}
\author{E.~Fiorini}\affiliation{\mib}
\author{G.Frossati}\affiliation{\ul}
\author{E.Guardincerri}\affiliation{\uge}
\author{A.~Giuliani}\affiliation{\co}
\author{P.~Gorla}\affiliation{\mib}\affiliation{\uz}
\author{E.E.~Haller}\affiliation{\lbnl}\affiliation{\ucb}
\author{J.~McDonald}\affiliation{\lbnl}
\author{E.B.~Norman}\affiliation{\lbnl}
\author{A.~Nucciotti}\affiliation{\mib}
\author{E.~Olivieri}\affiliation{\ufi}
\author{M.~Pallavicini}\affiliation{\uge}
\author{E.~Palmieri}\affiliation{\lnl}
\author{E.~Pasca}\affiliation{\ufi}
\author{M.~Pavan}\affiliation{\mib}
\author{M.~Pedretti}\affiliation{\co}
\author{G.~Pessina}\affiliation{\mib}
\author{S.~Pirro}\affiliation{\mib}
\author{E.~Previtali}\affiliation{\mib}
\author{L.~Risegari}\affiliation{\ufi}
\author{C.~Rosenfeld}, \affiliation{\usc}
\author{S.~Sangiorgio}\affiliation{\co}
\author{M.~Sisti}\affiliation{\mib}
\author{A.R.~Smith}\affiliation{\lbnl}
\author{L.~Torres}\affiliation{\mib}
\author{G.~Ventura}\affiliation{\ufi}


\begin{abstract}
We report the present results of CUORICINO a cryogenic experiment on neutrinoless 
double beta decay (DBD) of \tect consisting of  an array of 
62 crystals of \teod with a total active
mass of 40.7 kg.  The array is framed inside of a dilution refrigerator, heavily
shielded against environmental radioactivity and high-energy neutrons, 
and operated at a temperature of  \ca8 mK in the Gran Sasso Underground
Laboratory. Temperature pulses  induced by particle interacting in the crystals 
 are recorded and measured by means of Neutron Transmutation Doped
thermistors.   
The gain of each bolometer is stabilized with voltage pulses developed by a 
high stability pulse generator across heater resistors put in thermal 
contact with the absorber.
 The calibration is performed by means of two thoriated wires routinely inserted 
 in the set-up.
No evidence for a peak indicating neutrinoless DBD of  
\tect  is detected and a 90 \% C.L. lower  limit of 1.8\per 10$^{24}$ years is set 
for the lifetime of this process.
Taking largely into account the uncertainties in the theoretical values  of nuclear
matrix elements, this implies an upper boud  on the effective mass of the electron neutrino 
ranging from 0.2 to 1.1 eV. This sensitivity is similar to those of the $^{76}$Ge
experiments.
 
\end{abstract}

			      
\maketitle

Great interest  was stimulated  in recent  years in  neutrinoless double beta
decay (DBD) as a consequence of the observation of neutrino oscillations
\cite{Fukuda,Ahmad,Eguchi,Hagiwara,Ahn,Ahmed},             
proving that the differences between the squares  of the neutrino mass eigenvalues 
 is different from zero. This indicates that the mass m$_\nu$  of at 
least one neutrino is finite, but does not allow the determination of its absolute 
value.

The value of the sum of the  masses of the neutrinos of the three flavors has 
been constrained to values from 0.7 to 1.7 eV from the  WMAP full sky microwave map together 
with the survey of the 2dF galaxy redshift \cite{Barger,Hannestad,Tegmark,Spergel,Crotty}. 
A claim for a non zero value of 0.64 eV has also been proposed~\cite{Allen}. 
These values are more constraining than upper limits of 2.2 eV for  $m_\nu$ 
obtained so far in experiments on single beta decay, but they are strongly model
dependent and therefore less robust than laboratory measurements. Limits of \ca0.2 eV are 
expected in KATRIN experiment \cite{Lobashov}. 
If neutrinos are Majorana particles more
stringent constraints, or a positive value for the effective neutrino mass, can be 
obtained by neutrinoless DBD. In this lepton violating  process, a nucleus (A,Z) 
decays into (A,Z+2) with the emission of two electrons and no neutrino. This leads 
to a peak in the sum energy spectrum of the two electrons. The decay rate of this  process
would be proportional to the square of the effective neutrino mass  
\amnu, which can be expressed in terms of the elements of the neutrino mixing matrix as follows:
\begin{equation}\label{eq:meff}
\vert\langle m_{\nu} \rangle\vert \equiv \vert \vert
U_{e1}^L \vert ^2m_1 + \vert U_{e2}^L \vert ^2m_2 e^{i\phi _2 } +
\vert U_{e3}^L \vert ^2m_3 e^{i\phi _3 }\vert ,
\end{equation}

\noindent where $e^{i\phi _2 }$ and $e^{i\phi _3 }$ are the Majorana CP--phases (\pom1 for CP conservation), $m_{1,2,3} $ are the Majorana neutrino mass eigenvalues and  U$^L_{ej}$ are the coefficients of the Pontecorvo-Maki-Nakagawa-Sakata (PMNS) neutrino mixing matrix, determined from neutrino oscillation data. 
Recent global analyses of all oscillation experiments ~\cite{Feruglio,Ferugliobis,Joaquim,
Giunti,Pascoli,Pascolibis,Bahcall,Bahcallbis,Murayama,Suhonen} yield on average:

\begin{eqnarray}\label{eq:meffexp} 
\vert\langle m_{\nu} \rangle\vert = \vert (0.70\pm0.03) m_1 + (0.30\pm 0.03) m_2 e^{i\phi _2 } + \nonumber \\
+(<0.05) m_3 e^{i\phi _3 }\vert 
\end{eqnarray}

It should be stressed that neutrino oscillation experiments can only yield neutrino mass eigenvalue differences squared, and imply two possible patterns, or hierarchies, the normal: m$_1$\dca m$_2 << $m$_3$, and the inverted hierarchy: m$_1 << $m$_2$ \dca m$_3$. The mass parameter measured in solar oscillation experiments, $\delta_{solar}$, is m$_2^2$--m$_1^2$  in the normal hierarchy case and  m$_3^2$--m$_2^2$ in the inverted case. That measured in atmospheric neutrino experiments,  $\delta_{atm}$, is then approximately  m$_3^2$--m$_1^2$ in both cases. If we neglect  U$^L_{e3}$, and also note that experimentally, $\delta_{solar}<< \delta_{atm}$, two useful approximate expressions for $\vert\langle m_{\nu} \rangle\vert$  result:

\begin{equation}\label{eq:meffnh} 
\vert\langle m_{\nu} \rangle\vert = m_1\vert 0.70 + 0.3 e^{i\phi _2 }(1 +
\delta_{solar}^2/m_1^2)\vert 
\end{equation}

for normal hierarchy and

\begin{equation}\label{eq:meffih} 
\vert\langle m_{\nu} \rangle\vert = \sqrt{m_1^2+\delta_{atm}^2}\vert 0.70 e^{i\phi _2 } + 0.3 e^{i\phi _3 }\vert 
\end{equation}

for inverted hierarchy.
If one uses the value, $\delta_{atm}$=2\per10$^{-3}$, equation~(\ref{eq:meffih}) implies that \amnu  could have a minimum value as large as 0.045 eV, which implies a minimum sensitivity acceptable for next generation experiments on neutrinoless DBD.

One should note that the rate for
neutrinoless DBD is 
proportional also to the square of the nuclear matrix elements whose calculations are
still quite uncertain. As a consequence it is imperative to search neutrinoless
DBD in different nuclei. This is also true because  a peak attributed to this 
process could in
principle be mimicked by a radioactive line. Only the discovery of  peaks at
the different energies expected for neutrinoless DBD in two or more
candidate nuclei would  definitely prove  the existence  of this process.
No evidence for neutrinoless DBD has been reported so far
\cite{Tretyak,Barabash,Elliott,Elliottbis}, with the exception of  the  claimed  
discovery of the decay of $^{76}$Ge reported   by a subset of the
Heidelberg-Moscow collaboration \cite{Klapdor}. This claim has been contested by
various authors \cite{Feruglio,Aalseth,Zdesenko} and also by other members of the
same Heidelberg-Moscow Collaboration \cite{Bakaliarov}. A new  analysis in
favor of the previous claim  has
however been published recently \cite{Klapdorbis,Klapdorter}.

Here we report new results on the neutrinoless DBD of \tect from the CUORICINO
experiment operating  in the Gran Sasso Underground
Laboratory at a depth of about 3500 m.w.e.\cite{Bettini}. This search, like the previous 
ones performed  in
the same laboratory,  is carried out with the cryogenic technique suggested for the
first time twenty years ago for searches for rare events \cite{Fiorini}.
Cryogenic thermal detectors
\cite{Twerenbold,Booth} are made by diamagnetic and dielectric crystals kept at
low temperature, where their heat capacity is proportional to the cube 
of the temperature itself. As a consequence, their heat capacity can become so
small  that even the tiny energy delivered to this "absorber" by particle interaction , can 
be detected  and measured by means of a suitable thermal sensor. Since the only requirement 
for these absorbers is  that they have reasonable thermal and mechanical
properties, cryogenic detectors offer a wide choice of candidate nuclei for
searches on DBD. The \tect  nucleus is an excellent candidate due to its
high transition energy (2528.8 \pom 1.3  keV) \cite{Dyck},  and especially to 
the unusually  large
isotopic abundance ( 33.8\%) \cite{Firestone} which makes the need
for enrichment less important. A preliminary report on the first part of 
this experiment was published earlier \cite{Arnaboldi}.

\begin{figure}[t!]
\includegraphics[width=0.6\linewidth]{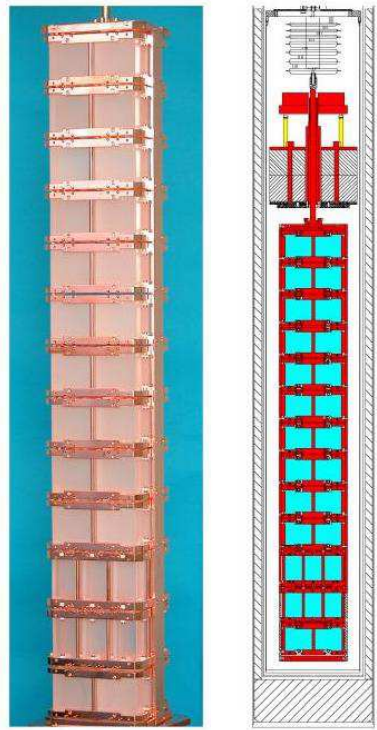}%
\caption{\label{fig:cuoricino} Scheme  of  CUORICINO.}
\end{figure}

CUORICINO (Fig.~\ref{fig:cuoricino}) is a tower of 13 planes containing 62 
 crystals  of \teod; 44 of them are cubes of 5 cm on a side while the dimensions of the 
 others are \magro. All crystals are made with natural paratellurite, apart from two \magro crystals, 
which are  enriched in $^{128}$Te and two others of the same size enriched in 
\tect, with 
isotopic abundance of 82.3 \% and   75 \%, respectively. The total mass of TeO$_2$ in CUORICINO
 is 40.7 kg, the largest  by more than an order of magnitude than any
cryogenic detector. More details
on the preparation of the crystals and on the mechanical structure of the 
array is reported elsewhere \cite{Arnaboldi}. 

In order to shield against the radioactive
contaminants from the materials of the refrigerator, a 10 cm layer  of Roman
lead,  with $^{210}$Pb activity of $<$4 mBq kg$^{-1}$ \cite{Alessandrello}, is inserted inside the 
cryostat immediately above the CUORICINO tower. A 1.2 cm lateral layer of the 
same lead is framed around the array to reduce  the activity of the
thermal shields. The cryostat is externally shielded by two layers of Lead of 10
cm minimal thickness. While the outer is made by common Lead, the inner one
has a $^{210}$Pb  activity  of (16 \pom 4) Bq kg$^{-1}$. An additional layer  of
~2 cm is provided by the electrolitic Copper of the thermal shields. The
background due to environmental neutrons is reduced by a layer of Borated
Polyethylene of 10 cm minimum thickness. The refrigerator operates inside a
Plexiglass anti-radon box flushed with clean N$_2$, and inside a Faraday cage to
reduce electromagnetic interference. 

Thermal pulses are recorded by means of Neutron Transmutation Doped (NTD) Ge
thermistors thermally coupled to each  crystal. 
Stabilization is performed  
by means of voltage pulses developed across heater resistors attached to each bolometer. 
The voltage pulses  are generated by  high stability pulse generators, designed and 
developed on purpose~\cite{Gianluigi1}. 
A tagging of these stabilizing signals is made by the acquisition system. 
The detector baseline is stabilized 
with a dedicated circuit with a precision of better than about 0.5 KeV/day 
on the average~\cite{Gianluigi2} 
between the successive refilling of liquid helium of the main reservoir.

The front-end electronics for all the \magro and
for 20 of the \ciccio detectors are mantained  at room temperature. In the so
called $\textit{cold electronics}$ , applied to the remaining 24 crystals, the
preamplifier is located in a box at \ca 100 K 
near the detector to reduce the noise due to microphonics \cite{Gianluigi3}, which would be very 
dangerous when searching for
WIMPS. More details on read-out electronics and DAQ are reported in
\cite{Arnaboldi}.

CUORICINO is operated at a temperature of \ca 8 mK with a spread of
\ca 1 mK. A routine energy calibration is performed before and after each
sub-run, which lasts about two weeks, by exposing the array to two thoriated 
tungsten wires inserted in
immediate contact with the refrigerator. All data, in which the average difference
between the initial and final calibration is larger than the experimental error
in the evaluation of the peak position were discarded. 

During the first cool down, 12 of the  \ciccio   and one of the 
\magro crystals were lost, due to the disconnections at the level of the
thermalisation stages  which allow the transmission of the electric signals from the
detectors to room temperature \cite{Arnaboldi}. Since the active mass was of
 \ca 30 kg, and the energy
resolution was excellent, data collection was continued  for a few months before
warming up the array. The problem has now been fully solved and the detector was
cooled down with 
only 2 of the 13 detectors still disconnected.
The data presented here come from the first run and about 3 months of the second run.
The total statistics corresponds to an effective exposure of 10.85 kg \per year.

\begin{figure}[t!]
\includegraphics[width=0.9\linewidth]{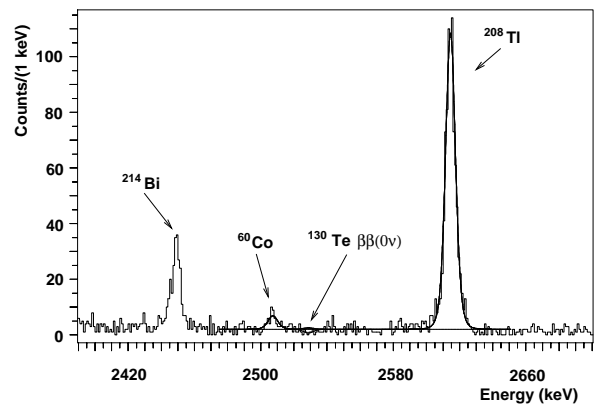}%
\caption{\label{fig:Spectrum} Spectrum of the sum of the two electron energies in
the region of neutrinoless DBD}
\end{figure}

The sum of the spectra of the \ciccio and \magro crystals in the 
region of the neutrinoless DBD is shown in Fig.~\ref{fig:Spectrum}. One can clearly see the
peaks at 2447 and 2615 keV from the decays of   $^{214}$Bi and $^{208}$Tl, plus a small peak
at 2505 keV due to the sum of the two $\gamma$ lines of  $^{60}$Co.  
The background at the energy of neutrinoless DBD is of 0.18 \pom 0.01 
counts kg$^{-1}$ keV$^{-1}$ y$^{-1}$.

 No evidence is found  for a peak at ~ 2529 keV, the energy
expected for neutrinoless DBD of $^{130}$Te. By applying a maximum likelihood
procedure \cite{Baker,Barnett}, we obtain a 90\% C.L. lower limit of 
1.8 \per 10$^{24}$ years on the lifetime for this decay. The unified approach of
G.I.Feldman and R.D.Cousins \cite{Feldman,Groom} leads to a similar result. 
The upper bounds on the effective mass of the electron neutrino that can be extracted from our
result depend strongly on the values adopted for the nuclear matrix elements. As
in our previous paper \cite{Arnaboldi} we considered all theoretical calculations 
\cite{Suhonen,Tretyak,Elliott,Elliottbis} apart from those based on the shell model 
which is not considered as
valid for heavy nuclei \cite{Faessler}, in particular for DBD of 
\tect \cite{Elliottbis}. We have also not considered the 
calculation by Rodin et al \cite{Rodin} based on the evaluation of the 
particle-particle interaction strenght from the corresponding two neutrino DBD 
lifetime. The evaluation based on single beta decay, which could be preferable
 \cite{Elliottbis,Suhonenbis} is not available for $^{130}$Te. The rates for  
two neutrino DBD of this nucleus based on geochemical experiments are however 
uncertaint \cite{Tretyak,Elliott,Elliottbis}. We have therefore adopted 
 \cite{Rodin} those based on a direct  experiment  \cite{Arnaboldibis}. 
  
Taking into account the above mentioned uncertainties, our lower limit leads to a
constraint on the effective mass of the electron neutrino ranging from 0.2 to 1.1 eV, and
partially covers the mass range of 0.1 to 0.9 eV  indicated by H.V. Klapdor-Kleingrothaus et al.~
\cite{Klapdorter}.

CUORICINO is a first step towards the realization of CUORE (Cryogenic
Underground Observatory for Rare Events). It would be  an array made by 19 towers, each
similar to CUORICINO, with  988 cubic crystals of \teod,
5 cm on a side, and a total active mass of 741 kg. 
The expected sensitivity on \amnu of this experiment is of the order of 30 meV, just below 
the above cited value of 45 meV favoured by current oscillation experiments for the inverted hierarchy.
CUORE has  already been approved by the Gran Sasso Scientific Committee and by 
the National Institute of Nuclear Physics (INFN).

Thanks are due to the Director and Staff of the Laboratori Nazionali del Gran
Sasso and to the technical staffs of our Laboratories.
This experiment has been partially supported by the Commission of European
Communities under contract HPRN-CT-2002-00322 , by the US Department of Energy
under contract number DE-AC03-76 SF 98 and by the National Science
Foundation.

During the run of this experiment in preparing which he so vitally contributed ,
our colleague Angel Morales , leader of the Zaragoza group, passed away. It was
a great loss for  science and a great personal loss for all of us.

\end{document}